\documentclass[preprint,secnumarabic,amssymb, nobibnotes, aps, prb]{revtex4-1}

\usepackage{color}
\usepackage{epsfig, graphics, graphicx, subfigure, wrapfig, lipsum, amsmath}
\usepackage{verbatim}
\usepackage{lipsum} 
\usepackage{dcolumn}
\usepackage{bm}
\usepackage{amsmath, braket}
\usepackage{array}
\usepackage{xr}
\usepackage{longtable}
\usepackage{tabularray, hyperref}

\newcommand{\roml}[1]{\lowercase\expandafter{\romannumeral #1\relax}}
\newcommand{\romu}[1]{\uppercase\expandafter{\romannumeral #1\relax}}

\begin{document}

\title{Higher-order thermal transport theory for phonon thermal transport in semiconductors using  lattice dynamics calculations and the Boltzmann transport equation}

\author{Ankit Jain}
\email{a\_jain@iitb.ac.in}
\affiliation{Mechanical Engineering Department, IIT Bombay, India}
\author{Yagyank Srivastava}
\affiliation{Mechanical Engineering Department, IIT Bombay, India}
\author{Amey G. Gokhale}
\affiliation{Mechanical Engineering Department, IIT Bombay, India}
\author{Nidheesh Virakante}
\affiliation{Mechanical Engineering Department, IIT Bombay, India}
\author{Hardik L. Kagdada}
\affiliation{Mechanical Engineering Department, IIT Bombay, India}
\date{\today}%

\begin{abstract}
The phonon thermal conductivity of semiconducting periodic solids can be obtained using the lattice dynamics calculations along with the Boltzmann transport equation  and with input from density functional theory calculations. These calculations have resulted in an excellent agreement with experiments without requiring any fitting parameters. However, over the last decade, many material systems have been identified where the lowest level lattice dynamics theory, which is based on the relaxation time approximation solution of the Boltzmann transport equation and considers potential energy surface sampling around the static equilibrium positions of atoms with only three-phonon scatterings, is proved insufficient in describing the thermal transport physics. 

In this article, we review these higher-order developments in the lattice dynamics theory to describe thermal transport in periodic semiconducting solids. We start with a brief discussion of the lowest-order theory and discuss its limitations along with proposed developments to address these limitations. We discuss prominent success cases of these higher-order developments and present our recommendations on their use for various material systems. Considering that many of these higher-order developments are computationally more demanding compared to the lowest-order theory, we also discussed data-driven approaches to accelerate these calculations. This review article is intended to serve as a reference for both novice and experienced researchers in this field. 
\end{abstract}

\maketitle

\section{Introduction}\label{sec_introduction}
The understanding of phonon thermal transport physics in semiconductors is crucial for applications such as thermoelectric energy conversion, heat dissipation, and memory storage devices  \cite{clarke2003,dames2005,lindsay2018survey}. The thermal transport in non-metallic crystalline materials (periodic semiconducting and insulating solids) is predominantly through atomic vibrations, i.e., phonons \cite{ziman1960, reissland1973}. Traditionally, the thermal design of semiconducting solids is led by resource-consuming experiments based on a trial-and-error approach guided by simple physics principles such as the presence of heavy atomic species and complex material unitcell \cite{slack1973}.  However, with recent advances in computational resources, it is now possible to predict the phonon thermal conductivity of materials without requiring any experimental fitting parameters \cite{esfarjani2008, esfarjani2011, lindsay2018survey, mcgaughey2019phonon, veeravenkata2021}; thus paving the way for computational discovery of materials with targeted thermal properties.

Lattice dynamics calculations and molecular dynamics simulations are two commonly employed tools for investigating thermal transport in solids \cite{mcgaughey2004a, turney2009a}. In lattice dynamics calculations, the atoms are treated as a coupled spring-mass system, and the phonon modes are obtained as eigenmodes of the coupled system with linear/harmonic springs.\cite{kittel2018introduction} The non-linearity/anharmonicity is then introduced as a perturbation to harmonic modes \cite{reissland1973}. This approach is ideal for perfectly periodic systems, and any deviation from this is handled via the perturbative treatment. The lattice dynamics approach provides a bottom-up mode-level understanding of the thermal transport physics, and the thermal conductivity is obtained by summing over the contributions of all modes.  In contrast, in molecular dynamics simulations, the atoms are allowed to move in real-time, and the time-evolution of atomic trajectories is obtained by using Newton's laws. These atomic trajectories are then used to obtain the thermal conductivity. Unlike lattice dynamics, the molecular dynamics simulations can account for interaction anharmonicity to the highest order, and the inclusion of lattice imperfections is trivial in these simulations. While these simulations treat atoms as classical particles (occupation governed by the Boltzmann distribution), methodologies to account for quantum corrections are also explored in the literature \cite{turney2009assessing}.

The thermal conductivity obtained from lattice dynamics calculations and molecular dynamics simulations depends on interatomic interactions \cite{esfarjani2008, esfarjani2011, mcgaughey2019phonon}. Consequently, the applicability of these approaches was originally limited to qualitative/trend predictions as the interactions were obtained from empirical forcefields \cite{mingo2003, lindsay2008, lindsay2010}, fitted to reproduce one or more experimentally observed properties of the given material \cite{brenner1990empirical, tersoff1988empirical, stillinger1985}. With the availability of ab-initio-based density functional theory (DFT) driven interatomic interactions, it is now possible to obtain the actual thermal conductivity of materials from these approaches, directly comparable with experiments \cite{broido2007, esfarjani2008, esfarjani2011, lindsay2018survey, mcgaughey2019phonon}. The lattice dynamics calculations have an advantage in this regard as they require interatomic force constant evaluation only once, while molecular dynamics simulations need interatomic interactions at each of the millions of simulation timesteps. Despite this advantage, the computational cost of interatomic force constants evaluation from DFT is high at several thousand to tens of thousands cpu-hours \cite{esfarjani2008} and approximate data-driven methodologies, such as based on machine learning, are being developed to reduce the computational cost \cite{guo2023, guo2023b, srivastava2023, srivastava2024, Srivastava2024b}.  

On top of interatomic interactions, the accuracy of results obtained from the lattice dynamics calculations also depend on several other factors such as (not limited to) (\roml{1})  methodology of interatomic force constant extraction (finite-difference, compressive-sensing, etc) \cite{lindsay2013b,  zhou2014lattice, carrete2016physically, xie2017effect}, (\roml{2}) phonon-phonon interactions (three-phonon, four-phonon scatterings, etc) \cite{feng2017four, feng2018, ravichandran2020}, (\roml{3}) phonon eigenmode renormalization (to account for anharmonicity) \cite{ravichandran2018unified}, and (\roml{4})  employed thermal transport theory [Boltzmann transport equation (BTE) based particle-like thermal transport or coherent-based wave-like transport] \cite{simoncelli2019, xia2020, jain2020}.

This tutorial-cum-review is focused on lattice dynamics based approach for phonon thermal conductivity prediction of periodic solids. We will begin with a brief discussion of basic lattice dynamics theory and its success stories.  Next, we will discuss limitations of this basic theory and the developments presented in the literature to address these limitations. We will next present computational challenges and associated data-driven approaches to address computational challenges. Lastly, we will discuss our recommendations on the use of lattice dynamics approach for different material systems. We note that, while for completeness, we present a brief discussion of basic lattice dynamics theory and its success cases, the readers are referred to our previous tutorial for an in-depth discussion of basic lattice dynamics theory \cite{mcgaughey2019phonon}.

\section{Phonon Thermal Conductivity}
\label{sec_BTE}
The phonon thermal transport theory is discussed in detail in Refs. \cite{ ziman1960, reissland1973, srivastava1990, dove1993}. The phonons at wavevector $\boldsymbol{q}$ are defined as eigenmodes (3$N$  modes, enumerated by mode-index $\nu$, where $N$ is the number of atoms in the unitcell) of the Dynamical matrix ${\boldsymbol{D}}_{{\boldsymbol{q}}}$ whose elements, $D_{\boldsymbol{q}}^{3(b-1)+\alpha, 3(b^{'}-1) + \beta}$, are given by:
\begin{equation}
\begin{split}
 \label{eqn_dynamical}
 D_{\boldsymbol{q}}^{3(b-1)+\alpha, 3(b^{'}-1) + \beta} = \\
 \frac{1}{\sqrt{m_b m_{b^{'}}}} \sum_{l^{'}} \Phi_{b0;b^{'}l^{'}}^{\alpha\beta} \exp{\{i[{\boldsymbol{q}}.( {\boldsymbol{r}}_{b^{'}l^{'}}  -  {\boldsymbol{r}}_{b0}  )] \}},
 \end{split}
\end{equation}
where the summation is over all unit-cells in the lattice, $m_b$ is the mass of atom $b$ in the unit-cell,  $\boldsymbol{r}_{bl}$ is the position vector of atom $b$ in the $l^{th}$ unit-cell, $\alpha$, $\beta$ $\in (x,y,z)$ denotes cartesian directions, and $\Phi_{ij}^{\alpha\beta}$ is the real-space ($ij, \alpha\beta$)-element of the harmonic force constant matrix $\boldsymbol{\Phi}$. The phonon frequencies, $\omega_{{\boldsymbol{q}}\nu}$, are obtained from the diagonalization of the dynamical matrix as:
\begin{equation}
 \label{eqn_eigen}
 \omega_{{\boldsymbol{q}}\nu}^{2} {\boldsymbol{e}}_{{\boldsymbol{q}}\nu} = {\boldsymbol{D}}_{{\boldsymbol{q}}} \cdot {\boldsymbol{e}}_{{\boldsymbol{q}}\nu},
\end{equation}
where ${\boldsymbol{e}}_{{\boldsymbol{q}}\nu}$ is the phonon eigenvector. In the lattice dynamics approach, the phonon thermal conductivity is obtained from the steady-state analysis of the BTE:
\begin{equation}
\label{eqn_BTE}
{\boldsymbol{v}_{{\boldsymbol{q}}\nu}} \cdot \nabla {n_{{\boldsymbol{q}}\nu}} = \left( \frac{\partial n}{\partial t} \right)_{coll},
\end{equation}
where the left-hand side represents phonon drift with ${\boldsymbol{v}_{{\boldsymbol{q}}\nu}}$,  $n_{{\boldsymbol{q}}\nu}$ as phonon group velocity and population, and the right side represents phonon scattering via different collision processes. The phonon group velocities are obtained as:
\begin{equation}
\label{eqn_vg}
{\boldsymbol{v}_{{\boldsymbol{q}}\nu}} = \frac{\partial \omega_{\boldsymbol{q}\nu}}{\partial \boldsymbol{q}}.
\end{equation}
The BTE (Eqn.~\ref{eqn_BTE}) governs phonon dynamics for all phonon modes in the crystal, and the equations for different modes are coupled through the collision term, requiring population of all phonon modes for its evaluation.  In equilibrium, i.e., in the absence of temperature gradient, the phonon population is given by the Bose-Einstein distribution as $n^o_{{\boldsymbol{q}}\nu} = \frac{1}{e^x-1}$ with $x = \frac{\hbar\omega_{{\boldsymbol{q}}\nu}}{k_\text{B}T}$ and  $\hbar$, $k_\text{B}$, and $T$ representing the reduced Planck constant,  the Boltzmann constant, and the temperature.

The phonon thermal conductivity is defined from Fourier's law as:
\begin{equation}
\label{eqn_Fourier}
q^" = -k\nabla T,
\end{equation}
where the phonon heat flux $q^"$ is obtained by summing over all phonon modes as:
\begin{equation}
\label{eqn_heatFlux}
q^" = \frac{1}{V}\sum_{\boldsymbol{q},\nu} \hbar \omega_{\boldsymbol{q}\nu} {\boldsymbol{v}_{{\boldsymbol{q}}\nu}}  n_{{\boldsymbol{q}}\nu},
\end{equation}
with $V$ representing the crystal volume. By solving the linearized form of Eqn.~\ref{eqn_BTE}  and using Eqns.~\ref{eqn_Fourier} and \ref{eqn_heatFlux}, the phonon thermal conductivity in the $\alpha$ direction, $k_{\alpha}$, is obtained as:
\begin{equation}
\label{eqn_k}
k_{\alpha} = \sum_{{\boldsymbol{q}}} \sum_{\nu} c_{{\boldsymbol{q}}\nu} v_{{\boldsymbol{q}}\nu, \alpha}^{2} \tau_{{\boldsymbol{q}}\nu, \alpha}, 
\end{equation}
where  $c_{{\boldsymbol{q}}\nu}$ is the phonon specific heat:
\begin{equation}
\label{eqn_cph}
 c_{{\boldsymbol{q}}\nu} = \frac{\hbar\omega_{{\boldsymbol{q}}\nu}}{V} \frac{\partial n^{o}_{{\boldsymbol{q}}\nu}}{\partial T} = \frac{k_\text{B} x^2 e^x }{(e^x-1)^2},
 \end{equation}
 $v_{{\boldsymbol{q}}\nu,\alpha}$ is the $\alpha$ component of phonon group velocity vector ${\boldsymbol{v}_{{\boldsymbol{q}}\nu}}$, and $\tau_{{\boldsymbol{q}}\nu, \alpha}$ is the phonon scattering time.

\section{Lowest-order Theory}
\label{sec_lowestTheory}
\subsection{Collision Term aka Phonon Scattering}
\label{sec_RTA}
The phonons obtained from Eqns.~\ref{eqn_dynamical}  and \ref{eqn_eigen} are defined for a perfectly periodic harmonic crystal, and any deviation from this perfect picture requires explicit perturbative treatment in the form of scattering mechanisms. Some of the intrinsic scattering mechanisms for phonons are phonon-phonon, phonon-electron, and phonon-isotope scatterings. In semiconducting solids with few or no free electrons, the scattering of phonons from electrons is weak compared to other intrinsic scattering mechanisms \cite{ziman1960}. The phonon-phonon scattering is owing to anharmonic interactions, and phonon-isotope scattering is owing to the presence of isotopic nuclei in naturally occurring materials \cite{reissland1973}. While it is possible to avoid/reduce phonon-isotope scattering via the isotope enrichment of materials, the phonon-phonon scattering is unavoidable in all bulk materials, though it becomes weak compared to other mechanisms at low temperatures. 

\begin{figure}
\centering
\epsfbox{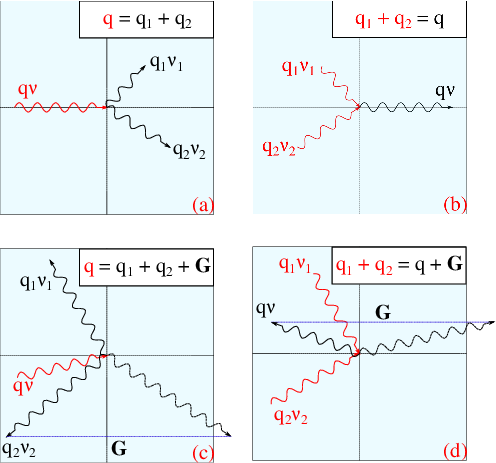}
\caption{\textbf {Three-phonon scattering processes.} In processes (a) and (c), one phonon splits into two phonons, while in (b) and (d), two phonons scatter to create a new phonon. 
The phonons continue to travel in the same direction after undergoing scattering via the \textit{Normal} scattering processes [(a) and (b)]. In (c) and (d), the scattering is via the \textit{Umklapp} scattering, and the resulting phonons (indicated by black dashed lines) are folded back into the first Brillouin zone by the reciprocal lattice vector $\boldsymbol{G}$  (blue dotted line). Phonons before and after the scattering events are shown in red and black, respectively.}
\label{fig_3ph}
\end{figure}

{\bf Phonon-phonon scattering:} The anharmonicity of interatomic interactions results in scattering of phonons from other phonons. At the lowest order  (i.e., first order in anharmonicity), the phonon-phonon scattering involves three phonons (three-phonon scattering), where either two phonons combine to create a new phonon or one phonon splits into two new phonons [Fig.~\ref{fig_3ph}]. The scattering rate of phonons due to three-phonon processes is given by:\cite{reissland1973}
\begin{equation}
 \begin{split}
 \label{eqn_3ph}
 \frac{1}{\tau_{\boldsymbol{q}\nu}^{3ph}} 
 =
  \sum_{{\boldsymbol{q}_{1}}\nu_{1}}
 \sum_{{\boldsymbol{q}_{2}}\nu_{2}}
 \bigg\{
 \Big\{
  {(n_{{\boldsymbol{q}_{1}\nu_{1}}}  - n_{{\boldsymbol{q}_{2}\nu_{2}}})}
 W^{+}
  \Big\} 
  + \\
  \frac{1}{2}
   \Big\{
  (n_{{\boldsymbol{q}_{1}\nu_{1}}} + n_{{\boldsymbol{q}_{2}\nu_{2}}} +  1)
  W^{-}
   \Big\}
   \bigg\},
 \end{split}
 \end{equation}
where the phonon scattering probabilities, $\boldsymbol{W}^\pm$,  are obtained from:
\begin{equation}
    \begin{split}
    \label{eqn_W_3ph}
W^{\pm}       
=
\frac{2\pi}{\hbar^2}
 \left|
 \Psi_{ {\boldsymbol{q}} (\pm{\boldsymbol{q}_{1}}) (-{\boldsymbol{q}_{2}}) }^{\nu \nu_{1} \nu_{2}}
 \right|^2
  \delta({\omega_{{{\boldsymbol{q}}_{}}\nu_{}} \pm \omega_{{{\boldsymbol{q}}_{1}}\nu_{1}} - \omega_{{{\boldsymbol{q}}_{2}}\nu_{2}}}),
    \end{split}
\end{equation}
with $\Psi_{ {\boldsymbol{q}} {\boldsymbol{q}_{1}} {\boldsymbol{q}_{2}} }^{\nu \nu_{1} \nu_{2}}$ denoting the Fourier transformation of real-space cubic force constants ($\Psi^{\alpha\beta\gamma}_{bl;b^{'} l^{'};b^{''} l^{''}}$):
\begin{equation}
    \begin{split}
        \label{eqn_cubic_Fourier}
\Psi_{ {\boldsymbol{q}} {\boldsymbol{q}_{1}} {\boldsymbol{q}_{2}} }^{\nu \nu_{1} \nu_{2}}
=
 \Psi_{ {\boldsymbol{q}} {\boldsymbol{q}^{'}} {\boldsymbol{q}^{''}} }^{\nu \nu^{'} \nu^{''}} = 
 N
 {\left(\frac{\hbar}{2N}\right)}^{\frac{3}{2}}
 \sum_{b} \sum_{b^{'} l^{'}}
\sum_{b^{''} l^{''}} 
\sum_{\alpha\beta\gamma} 
\Psi^{\alpha\beta\gamma}_{bl;b^{'} l^{'};b^{''} l^{''}} \\
\times 
\frac{
{{\tilde{e}}_{b,\boldsymbol{q}\nu}^{\alpha}}  
{{\tilde{e}}_{b^{'},{\boldsymbol{q}}^{'} \nu^{'}}^{\beta}} 
{{\tilde{e}}_{b^{''},{\boldsymbol{q}}^{''} \nu^{''}}^{\gamma}} }{\sqrt{ 
{m_b \omega_{\boldsymbol{q}\nu}}  
{m_{b^{'}} \omega_{{\boldsymbol{q}}^{'}\nu^{'}}}   
{m_{b^{''}} \omega_{{\boldsymbol{q}}^{''}\nu^{''}}}   }}  
e^{[i( {{\boldsymbol{q}}^{'}}  \cdot{\boldsymbol{r}}_{0l^{'}} 
+    {{\boldsymbol{q}}^{''}}  \cdot{\boldsymbol{r}}_{0l^{''}} )]}.
    \end{split}
\end{equation}
The summation in Eqn.~\ref{eqn_3ph} is performed over phonon wavevectors satisfying crystal momentum conservation, i.e., $\boldsymbol{q} + \boldsymbol{q_1} + \boldsymbol{q_2} = \boldsymbol{G}$, with $G$ being the reciprocal space lattice vector, and $\delta$ in Eqn.~\ref{eqn_W_3ph} is the Dirac delta function ensuring energy conservation.

{\bf Phonon-isotope Scattering:} The phonon-isotope scattering rate is obtainable using the Tamura theory by employing the mass disorder as:\cite{tamura1983}
\begin{equation}
    \begin{split}
    \label{eqn_rta_iso}
 \frac{1}{\tau_{\boldsymbol{q}\nu}^{iso}}
 =
 \sum_{{\boldsymbol{q}_{1}}\nu_{1}}
  n_{\boldsymbol{q}\nu}
 (n_{{\boldsymbol{q}}_{1}\nu_{1}}+1)
 W^{iso},
    \end{split}
\end{equation}
where $ W^{iso}$ represents the scattering probability matrix for isotope scattering given by:
\begin{equation}
    \begin{split}
    \label{eqn_W_iso}
 W^{iso} =    
\frac{\pi}{2N}
  \omega_{{\boldsymbol{q}}\nu} \omega_{{\boldsymbol{q}}_{1}\nu_{1}} 
 \sum_{b} g_2(b) 
 \left|
 {\boldsymbol{e}}^{*}_{b,\boldsymbol{q}\nu} 
 \cdot
 {\boldsymbol{e}}^{*}_{b,{\boldsymbol{q}}_{1}\nu_{1}}
 \right|^2 
 \times \\
 \delta 
 \left(
  \omega_{{\boldsymbol{q}}\nu} - \omega_{{\boldsymbol{q}}_{1}\nu_{1}} 
 \right),
    \end{split}
\end{equation}
with $g_2(b)$ as the mass variance parameter:
\begin{equation}
 \label{eqn_iso_g}
 g_2(b) = \sum_s f_s(b) \left( 1 - \frac{m_{b,s}}{{\overline{m}}_b}\right)^2,
\end{equation}
 $f_s(b)$ is the mass-fraction of isotope $s$ of atom $b$ with mass $m_{b,s}$ and ${{\overline{m}}_b}$ is the average mass of atom $b$.

\subsection{Relaxation Time Approximation}
\label{sec_RTA}
The evaluation of phonon scattering rates via three-phonon scattering processes using Eqn.~\ref{eqn_3ph} requires population of all phonon modes in the system ${(n_{{\boldsymbol{q}_{1}\nu_{1}}},n_{{\boldsymbol{q}_{2}\nu_{2}}})}$. These phonon populations are to be obtained by solving Eqn.~\ref{eqn_BTE}, thus coupling differential equations of all phonon modes (phonon thermal conductivity evaluation typically requires $\sim$ 10,000 phonon modes). A common assumption to decouple these equations is to approximate phonon populations by their equilibrium populations (Bose-Einstein distribution) while evaluating phonon scattering rates using Eqn.~\ref{eqn_3ph}. This assumption is called Relaxation Time Approximation (RTA), and the phonon lifetimes obtained using this assumption are referred to as phonon relaxation times (opposed to phonon scattering times) \cite{ziman1960, ward2009ab, esfarjani2011}.

\subsection{Interatomic Force Constants}
\label{sec_finiteDifference}

The evaluation of phonon thermal conductivity via Eqn.~\ref{eqn_k} requires phonon heat capacity, group velocity, and scattering rates. The evaluation of phonon heat capacity and group velocity require harmonic force constants ($\Phi_{ij}^{\alpha\beta}$) while the evaluation of phonon scattering rates requires cubic force constants ($\Psi^{\alpha\beta\gamma}_{ijk}$) [and potentially higher-order (quartic force constants,  $\Xi_{ijkl}^{\alpha\beta\gamma\delta}$)]. The harmonic and cubic force constants are second and third derivatives of the system potential energy defined as:\cite{esfarjani2008}
\begin{equation}
\Phi_{ij}^{\alpha\beta} = \frac{\partial^2 U}{\partial u_i^{\alpha} \partial u_j^{\beta}} = -\frac{\partial F_i^{\alpha}}{\partial u_j^{\beta}},
\end{equation}
\begin{equation}
\Psi_{ijk}^{\alpha\beta\gamma} = \frac{\partial^3 U}{\partial u_i^{\alpha} \partial u_j^{\beta} \partial u_k^{\gamma}} = -\frac{\partial^2 F_i^{\alpha}}{\partial u_j^{\beta} \partial u_k^{\gamma}},
\end{equation}

using the Taylor series expansion of the system's potential energy:
\begin{equation}
\begin{split}
\label{eqn_PE}
 U = U_0 + 
 \displaystyle\sum_{i} {\Pi_{i}^{\alpha} u_i^{\alpha}} +
\frac{1}{2!}\displaystyle\sum_{ij} {\Phi_{ij}^{\alpha\beta}u_i^{\alpha}u_j^{\beta}} + \\
\frac{1}{3!}\displaystyle\sum_{ijk} {\Psi_{ijk}^{\alpha\beta\gamma} u_i^{\alpha}u_j^{\beta}u_k^{\gamma}} +
\\
\frac{1}{4!}\displaystyle\sum_{ijkl} {\Xi_{ijkl}^{\alpha\beta\gamma\delta} u_i^{\alpha}u_j^{\beta}u_k^{\gamma}u_l^{\delta}} +
O\left({u^5}\right),
\end{split}
\end{equation}
where $u_i^{\alpha}$ is the displacement of atom $i$ in the $\alpha$-direction from its equilibrium position, $F_i^{\alpha}$ is the force on atom $i$ in the $\alpha$ direction,  $U_o$ is arbitrary reference, and $\Pi^{\alpha}_{i}$ is the first-derivative of potential energy representing forces on atoms (zero when the above expression is written for equilibrium/relaxed positions of atoms). In the simplest approach, the harmonic and cubic force constants are obtainable from finite difference of atomic forces by displacing one or more atoms around the equilibrium positions \cite{mcgaughey2019phonon}. Alternatively, all atoms could be displaced from the equilibrium positions by a small random amount, and the force constants could be obtained via Taylor-series fitting of the generated force-displacement dataset \cite{esfarjani2008, zhou2014lattice}. Both of these approaches sample the material potential energy surface in the vicinity of relaxed positions of atoms, i.e., when atoms are populated without thermal displacements at 0 K temperature.

\begin{figure}
\centering
\epsfbox{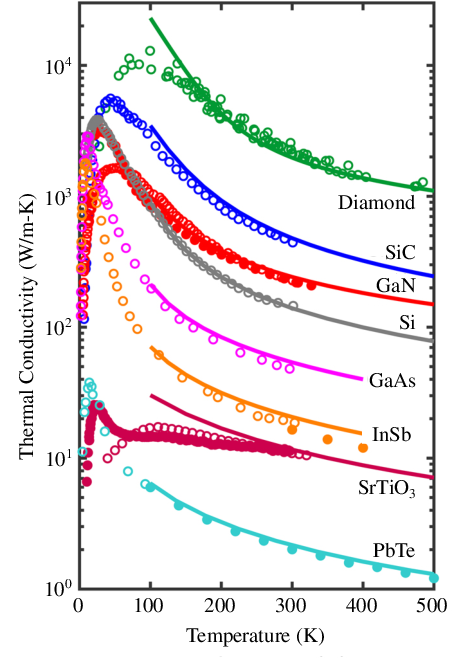}
\caption{\textbf{Thermal conductivity from the lowest-order theory.} Comparison of lattice thermal conductivities at different temperatures computed from first-principles lattice dynamics approach with the lowest-order thermal transport theory (solid lines) against experimental data (circles). The experimental and computational data are from Refs.~\cite{onn1992some,olson1993thermal,wei1993,berman1975nitrogen,morelli2002estimation,slack2002some,jezowski2003thermal,inyushkin2004,holland1964phonon,maycock1967thermal,yu2008thermal,buscaglia2014effect,greig1960thermoelectricity,ravich2013semiconducting} and ~\cite{ward2009ab,lindsay2013b,feng2015phonon,shiga2012microscopic} respectively. [Reprinted with permission from Ref. \cite{mcgaughey2019phonon}].}
\label{fig_LT}
\end{figure}

\subsection{Success Cases}
\label{sec_results_LT}
The first set of results on the prediction of phonon thermal conductivity using first-principles based interatomic force constants was reported by Esfarjani et al. for silicon in 2011 \cite{esfarjani2011}. The authors discussed the use of crystal symmetries and invariance constraints to reduce the required number of first-principles-based calculations \cite{esfarjani2008}. In particular, the authors showed that symmetries are critical for these calculations and can help in reducing the required number of calculations from 1146 to 5 for cubic force constant evaluations in silicon. The obtained phonon thermal conductivity was 166 W/m-K at room temperature, which was within 5\% of the experimentally measured value. Subsequently, many studies have focused on the effect of numerical/simulation parameters \cite{jain2015b} (such as supercell size \cite{lindsay2013b}, interaction cutoffs \cite{lindsay2013b}, and implementation of invariance constraints \cite{li2012}) on predicted thermal conductivity and prediction of thermal conductivity for diverse material systems spanning simple and compound semiconductors \cite{lindsay2013b}, bulk and planar materials \cite{jain2015strongly, veeravenkata2021}, naturally occurring and isotopically enriched materials \cite{lindsay2012}, and pristine and defective materials \cite{li2012}. We refer the readers to Ref.~\cite{lindsay2018survey, Togo2023} for a review of these studies. Overall, these studies resulted in an excellent agreement with experimental measurements, where possible, for instance, as shown in Fig.~\ref{fig_LT} for bulk crystalline materials with a varying range of thermal conductivities.

\section{Limitations of lowest-order Thermal Transport Theory}
\label{sec_limitations}

While the lowest-order theory is still widely used and results in a general agreement between prediction and experiments for a large class of materials, there are known shortcomings/fail-cases of this theory, which are discussed here.

\subsection{Full Solution of the BTE: Failure of the  RTA approximation}
\label{sec_failRTA}
The failure of the RTA solution of the BTE was reported even before the first ab-initio-based thermal conductivity prediction. This failure was reported by Broido et al.~\cite{ward2009ab} in 2009 for diamond based on an approach suggested by Omini et al.\cite{omini1996} in 1996. The authors found that the RTA solution results in an under-prediction of phonon thermal conductivity by 50\% for isotopically enriched diamond at 300 K. The authors discussed that this originates from resistive treatment of \textit{Normal} scattering processes by the RTA solution. As presented in the top panels of Fig.~\ref{fig_3ph}, the phonons undergoing scattering via \textit{Normal} processes continue to travel in the same direction after scattering and as such are non-resistive for thermal transport. For \textit{Umklapp} processes, however, the phonon travel direction reverses with scattering, resulting in resistance to heat transfer.  The RTA solution does not distinguish between \textit{Normal} vs \textit{Umklapp} processes and considers both as resistive. 

\begin{figure*}
\centering
\epsfbox{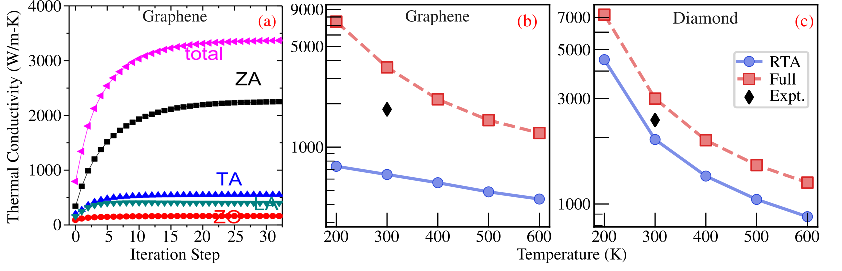}
\caption{\textbf{Full solution of the BTE.} The change in predicted thermal conductivity with full solution of linearized BTE. In (a), the contribution of different phonon modes towards the total thermal conductivity of graphene at 300 K is reported as a function of iteration number (Iteration-0 corresponds to the RTA solution). In (b) and (c), the temperature-dependent thermal conductivities of graphene and diamond are reported from the RTA and the full solution of the BTE. The experimental datapoints in (b) and (c) are from Refs.~\cite{xu2014length} and \cite{wei1993} respectively. [Reprinted with permission from Refs.~\cite{feng2018,ward2009ab}] }
\label{fig_iter}
\end{figure*}

This assumption can be lifted by considering a linearized form of the BTE (Eqn.~\ref{eqn_BTE}) for all phonon modes and by solving the coupled equations directly via matrix inversion. This approach is more involved from an implementation perspective and requires a large memory. This direct approach is employed by Chaput\cite{chaput2013direct} for obtaining frequency-dependent dynamic thermal conductivity of C, Si, and $\text{Mg}_2\text{Si}$. Alternatively, the coupled linearized BTE can also be solved iteratively until convergence, starting with the Bose-Einstein population as the initial guess population for all modes \cite{omini1996}. The iterative approach is less involved and is more commonly employed due to smaller memory requirements \cite{jain2015strongly, ShengBTE_2014, jain2024}. The iterations are performed on scattering lifetimes \cite{omini1996, ward2009ab, ravichandran2018unified, feng2018}:
\begin{equation}
\begin{split}
    \label{eqn_iterate_bte}
    \tau_{\boldsymbol{q}\nu} = \tau_{\boldsymbol{q}\nu}^o 
    (1 + 
    \Delta_{\boldsymbol{q}\nu}^{3ph} +
    \Delta_{\boldsymbol{q}\nu}^{iso} 
    ),
    \end{split}
\end{equation}
where $\Delta_{\boldsymbol{q}\nu}^{3ph}$ and $\Delta_{\boldsymbol{q}\nu}^{iso}$ are defined as:
\begin{equation}
\begin{split}
\label{eqn_iterate_tau_3}
\Delta_{\boldsymbol{q}\nu}^{3ph}
= 
 \sum_{{\boldsymbol{q}_{1}}\nu_{1}}
 \sum_{{\boldsymbol{q}_{2}}\nu_{2}}
 \bigg\{
 \Big\{
  {(n_{{\boldsymbol{q}_{1}\nu_{1}}}  - n_{{\boldsymbol{q}_{2}\nu_{2}}})}
 W^{+} 
\\ \times 
 ( \tau_{\boldsymbol{q}_2\nu_2}\xi_{\boldsymbol{q}_{}\boldsymbol{q}_{2}}^{\nu_{}\nu_{2}} 
 -  \tau_{\boldsymbol{q}_1\nu_1}\xi_{\boldsymbol{q}_{}\boldsymbol{q}_{1}}^{\nu_{}\nu_{1}}  )
  \Big\} 
  +
  \frac{1}{2}
   \Big\{
  (n_{{\boldsymbol{q}_{1}\nu_{1}}} + n_{{\boldsymbol{q}_{2}\nu_{2}}} +  1)
  W^{-}
 \\ \times 
  ( \tau_{\boldsymbol{q}_2\nu_2}\xi_{\boldsymbol{q}_{}\boldsymbol{q}_{2}}^{\nu_{}\nu_{2}} 
 + \tau_{\boldsymbol{q}_1\nu_1}\xi_{\boldsymbol{q}_{}\boldsymbol{q}_{1}}^{\nu_{}\nu_{1}}  )
   \Big\}
   \bigg\},
\end{split}
\end{equation}
\begin{equation}
    \begin{split}
    \label{eqn_iterate_tau_iso}
 \Delta_{\boldsymbol{q}\nu}^{iso}
 =
 \sum_{{\boldsymbol{q}_{1}}\nu_{1}}
  n_{\boldsymbol{q}\nu}
 (n_{{\boldsymbol{q}}_{1}\nu_{1}}+1)
 W^{iso}
 \tau_{\boldsymbol{q}_1\nu_1}\xi_{\boldsymbol{q}_{}\boldsymbol{q}_{1}}^{\nu_{}\nu_{1}},
    \end{split}
\end{equation}
with
\begin{equation}
    \xi_{\boldsymbol{q}_{}\boldsymbol{q}_{'}}^{\nu_{}\nu_{'}} 
    = 
    \frac{
    \omega_{\boldsymbol{q}_{'}\nu_{'}}v_{\alpha, \boldsymbol{q}_{'}\nu_{'}}
    }{
    \omega_{\boldsymbol{q}_{}\nu_{}}v_{\alpha, \boldsymbol{q}_{}\nu_{}}
    },
\end{equation}
and $\tau^o_{\boldsymbol{q}\nu}$ is the total phonon scattering lifetime obtained from the RTA solution (relaxation time, Eqns.~\ref{eqn_3ph}, \ref{eqn_rta_iso}) as 
\begin{equation}
\frac{1}{\tau^o_{\boldsymbol{q}\nu}} = \frac{1}{\tau^{3ph}_{\boldsymbol{q}\nu}} + \frac{1}{\tau^{iso}_{\boldsymbol{q}\nu}}.
\end{equation}

As discussed above, the thermal conductivity obtained via the full solution of the BTE differs from that obtained using the RTA solution for materials/conditions resulting in the dominant scattering of phonons via the \textit{Normal} processes. This happens: (\roml{1}) at low temperatures when only close-to Gamma-point phonon modes in the Brillouin zone are active/populated \cite{cepellotti2015, lee2015, harish2023}, (\roml{2}) for stiff materials or materials made of light atoms such as those based on carbon \cite{ward2009ab}, where phonon dispersion is very steep and modes close-to Brillouin-zone center are only populated, and (\roml{3}) for two-dimensional materials with quadratic dispersion, resulting in a very large density of flexural/close-to-Gamma-point modes \cite{lindsay2010}. 

Based on first-principles calculations, the under-prediction of thermal conductivity due to the RTA solution is reported to be, at room temperature, 50\% for diamond \cite{ward2009ab}, 72\% for graphite \cite{fugallo2014thermal}, and less than 10\% for silicon \cite{ward2009ab}. For graphene, with stiff bonds, light carbon atoms, and a two-dimensional structure, this difference is more than of factor of 3-5 at 300 K using only three-phonon scattering \cite{lindsay2014}. For other two-dimensional materials, such as $\text{MoS}_2$ \cite{gu2014phonon, gokhale2021cross, gokhale2023cross}, phosphorene \cite{jain2015strongly}, biphenylene \cite{veeravenkata2021}, MXenes \cite{gholivand2019effect},  etc, the difference varies between 25-40\%.

It is worthwhile to mention that second-sound and hydrodynamic thermal transport phenomena are also closely related to the relative strength of \textit{Normal} scattering processes compared to \textit{Umklapp} processes \cite{cepellotti2015, lee2015}. As such, materials/conditions resulting in failure of the RTA solution of BTE are akin to second sound and hydrodynamic thermal transport \cite{cepellotti2015, lee2015, huberman2019, ding2022, harish2023}.

\subsection{Higher-order Phonon-phonon Scattering}
\label{sec_fourPhonon}
In the last decade, many materials/conditions have been identified where the lowest-order treatment of phonon-phonon scattering (via three-phonon scattering) is found to be insufficient in describing the thermal transport physics \cite{ravichandran2018unified, feng2017four, xia2018revisiting, kagdada2025anomalous}. These materials/conditions are when: (\roml{1}) material/conditions are strongly anharmonic as indicated by low thermal conductivity [such as at high temperatures (with respect to the Debye temperature of material) or Van der Wall bonded materials (across layers in planar materials)] \cite{jain2022, gokhale2023cross} or (\roml{2}) when three-phonon scattering is low to start with (for instance, for materials with frequency bandgaps in phonon dispersion or symmetry-forbidden selection rules) \cite{lindsay2013b, han2023thermal}. 

\begin{figure}
\centering
\epsfbox{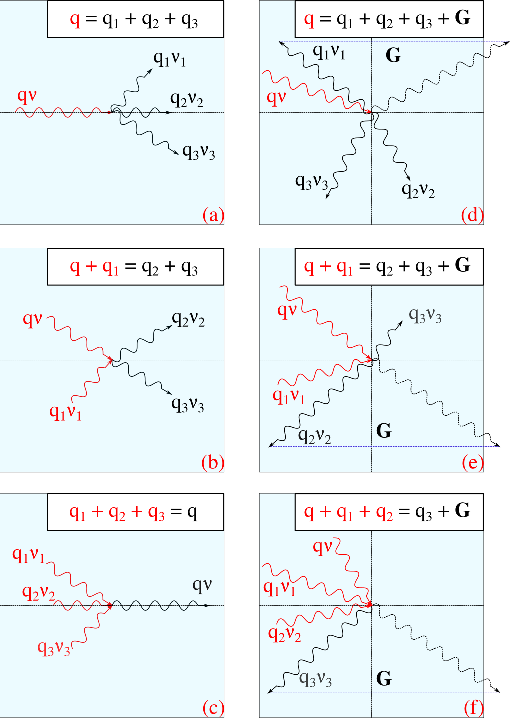}
\caption{\textbf{Four-phonon scattering processes}. In processes (a) and (d), one phonon splits into three new phonons. In processes (b) and (e), two phonons scatter together, exchanging their momentum and energy to create two new phonons. In processes (c) and (f), three phonons scatter to form a new phonon. Processes (a), (b), and (c) represent momentum-conserving \textit{Normal} scattering, while processes (d), (e), and (f) depict momentum-destroying \textit{Umklapp} scattering, where the resulting phonons (indicated by black dashed lines) are folded back into the first Brillouin zone by the reciprocal lattice vector $\boldsymbol{G}$  (blue dotted line). Phonons before and after the scattering events are shown in red and black, respectively.}
\label{fig_4ph}
\end{figure}

Similar to three-phonon processes, the four-phonon processes could take place through one phonon splitting to form three new phonons, two phonons getting scattered to form two new phonons, and three phonons getting scattered to create one new phonon, as shown in Fig.\ref{fig_4ph}. The scattering rate of phonons due to these processes can be obtained as \cite{feng2017four, feng2018, ravichandran2018unified, jain2020}:
\begin{equation}
 \begin{split}
  \label{eqn_4ph}
  \frac{1}{\tau_{\boldsymbol{q}\nu}^{4ph}}
 =
 \sum_{{\boldsymbol{q}_{1}}\nu_{1}}
 \sum_{{\boldsymbol{q}_{2}}\nu_{2}}
 \sum_{{\boldsymbol{q}_{3}}\nu_{3}}
 \bigg\{
 \frac{1}{6}
 \Big\{
 \frac{ 
 n_{{\boldsymbol{q}_{1}\nu_{1}}}  
 n_{{\boldsymbol{q}_{2}\nu_{2}}} 
 n_{{\boldsymbol{q}_{3}\nu_{3}}}  } 
 {
 n_{{\boldsymbol{q}_{}\nu_{1}}}
 }
 W^{--}
 \Big\}
 + \\
\frac{1}{2}
 \Big\{
 \frac{ 
 (n_{{\boldsymbol{q}_{1}\nu_{1}}}  + 1 )
 n_{{\boldsymbol{q}_{2}\nu_{2}}} 
 n_{{\boldsymbol{q}_{3}\nu_{3}}}  } 
 {
 n_{{\boldsymbol{q}_{}\nu_{1}}}
 }
  W^{+-}
 \Big\}
  +
  \\
  \frac{1}{2}
 \Big\{
 \frac{ 
 (n_{{\boldsymbol{q}_{1}\nu_{1}}}  + 1 )
 (n_{{\boldsymbol{q}_{2}\nu_{2}}} +1)
 n_{{\boldsymbol{q}_{3}\nu_{3}}}  } 
 {
 n_{{\boldsymbol{q}_{}\nu_{1}}}
 }
  W^{++}
 \Big\}
 \bigg\},
\end{split}
\end{equation}
where scattering probability matrix, $W^{\pm\pm}$, is given by: 
\begin{equation}
    \begin{split}
    \label{eqn_W_4ph}
W^{\pm\pm}       
=
\frac{2\pi}{\hbar^2}
 \left|
 \Xi_{ {\boldsymbol{q}} (\pm{\boldsymbol{q}_{1}}) (\pm{\boldsymbol{q}_{2}}) (-{\boldsymbol{q}_{2}}) }^{\nu \nu_{1} \nu_{2} \nu_{3}}
 \right|^2
 \times \\
  \delta({
  \omega_{{{\boldsymbol{q}}_{}}\nu_{}} \pm 
  \omega_{{{\boldsymbol{q}}_{1}}\nu_{1}} \pm 
  \omega_{{{\boldsymbol{q}}_{2}}\nu_{2}} -
  \omega_{{{\boldsymbol{q}}_{3}}\nu_{3}}
  }),
    \end{split}
\end{equation}
with $ \Xi_{ {\boldsymbol{q}} {\boldsymbol{q}_{1}} {\boldsymbol{q}_{2}} {\boldsymbol{q}_{2}} }^{\nu \nu_{1} \nu_{2} \nu_{3}}$ being the Fourier transform of the quartic force constants obtained as:
\begin{equation}
    \begin{split}
        \label{eqn_quartic_IFC}
\Xi_{ {\boldsymbol{q}} {\boldsymbol{q}_{1}} {\boldsymbol{q}_{2}} {\boldsymbol{q}_{3}} }^{\nu \nu_{1} \nu_{2} \nu_{3} }
=
 \Xi_{ {\boldsymbol{q}} {\boldsymbol{q}^{'}} {\boldsymbol{q}^{''}}  {\boldsymbol{q}^{'''}} }^{\nu \nu^{'} \nu^{''} \nu^{'''} } = \\
 N
 {\left(\frac{\hbar}{2N}\right)}^{{2}}
 \sum_{b} \sum_{b^{'} l^{'}}
\sum_{b^{''} l^{''}} 
\sum_{b^{'''} l^{'''}} 
\sum_{\alpha\beta\gamma\delta} 
\Xi^{\alpha\beta\gamma\delta}_{bl;b^{'} l^{'};b^{''} l^{''};b^{'''} l^{'''}}
\times \\
\frac{
{{\tilde{e}}_{b,\boldsymbol{q}\nu}^{\alpha}}  
{{\tilde{e}}_{b^{'},{\boldsymbol{q}}^{'} \nu^{'}}^{\beta}} 
{{\tilde{e}}_{b^{''},{\boldsymbol{q}}^{''} \nu^{''}}^{\gamma}} 
{{\tilde{e}}_{b^{'''},{\boldsymbol{q}}^{'''} \nu^{'''}}^{\delta}} }
{\sqrt{ 
{m_b \omega_{\boldsymbol{q}\nu}}  
{m_{b^{'}} \omega_{{\boldsymbol{q}}^{'}\nu^{'}}}   
{m_{b^{''}} \omega_{{\boldsymbol{q}}^{''}\nu^{''}}}   
{m_{b^{'''}} \omega_{{\boldsymbol{q}}^{'''}\nu^{'''}}} 
}}  
\times \\
e^{[i( {{\boldsymbol{q}}^{'}}  \cdot{\boldsymbol{r}}_{0l^{'}} 
+    {{\boldsymbol{q}}^{''}}  \cdot{\boldsymbol{r}}_{0l^{''}} 
+    {{\boldsymbol{q}}^{'''}}  \cdot{\boldsymbol{r}}_{0l^{'''}}   )]}.
    \end{split}
\end{equation}
The $\Xi^{\alpha\beta\gamma\delta}_{bl;b^{'} l^{'};b^{''} l^{''};b^{'''} l^{'''}}$ in Eqn.~\ref{eqn_quartic_IFC} represents quartic real-space force constants and the $\delta$ in Eqn.~\ref{eqn_W_4ph} is the Dirac delta function satisfying energy conservation. The summation in Eqn.~\ref{eqn_quartic_IFC} is carried out over all phonons satisfying $\boldsymbol{q} + \boldsymbol{q_1} + \boldsymbol{q_2} + \boldsymbol{q_3} = \boldsymbol{G}$. 


\begin{figure*}
\centering
\epsfbox{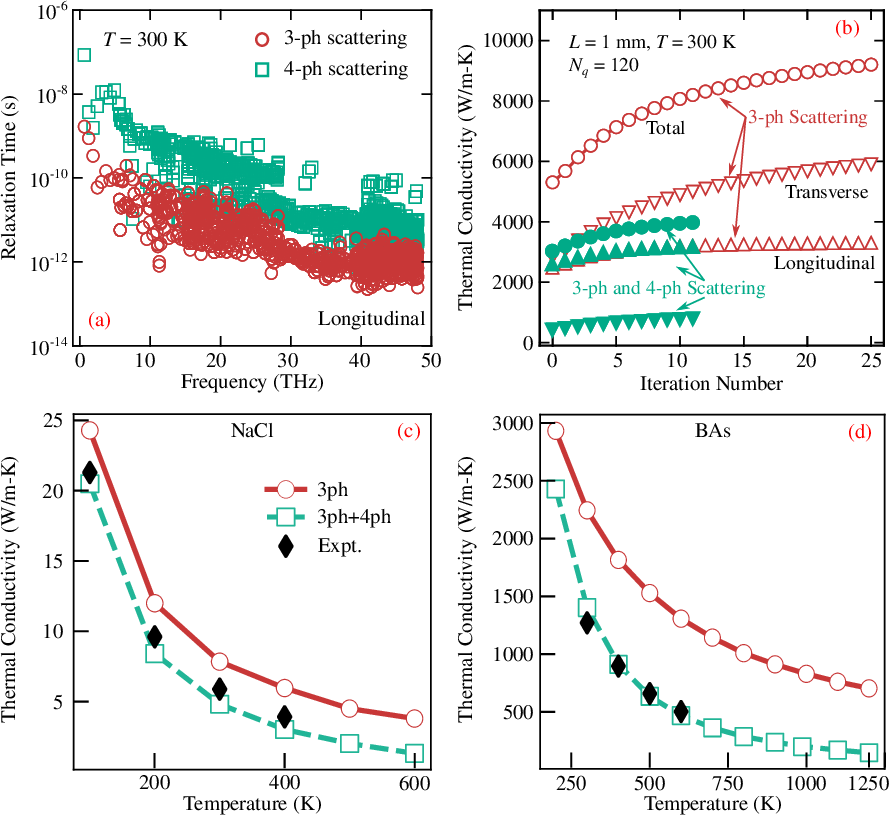}
\caption{\textbf{Effect of four-phonon scattering on phonon properties.} (a) The comparison of three-phonon and four-phonon scattering rates in single-walled carbon nanotubes and (b) The effect of four-phonon scattering on the predicted thermal conductivity of single-walled carbon nanotubes using the full solution of the BTE. [Reprinted with permission from Ref.~\cite{jain2024}]. The comparison of phonon thermal conductivity against experiments for (c) NaCl \cite{ravichandran2018unified} and (d) BAs \cite{feng2017four,kang2018experimental} with inclusion of four-phonon scattering.}
\label{fig_4ph}
\end{figure*}

The examples of materials where four-phonon scattering is significant owing to strong anharmonicity include NaCl \cite{ravichandran2018unified}, PbTe \cite{xia2018revisiting}, and $\text{Tl}_3\text{VSe}_4$ \cite{xia2020, jain2020}. For NaCl and PbTe, thermal conductivity predicted using 0 K IFCs and three-phonon scattering is also comparable with experimental measurements. However, this agreement arises due to a cancellation of errors from temperature-dependent phonon softening and the neglect of four-phonon scattering \cite{ravichandran2018unified, xia2018revisiting}. In $\text{Tl}_3\text{VSe}_4$, four-phonon scattering results in a factor of two reduction in predicted thermal conductivity \cite{jain2020}. Similarly, for cross-plane thermal transport in layered materials, the weak Van der Waals interactions render a pronounced effect of four-phonon scattering, with thermal conductivity reducing by more than 13\% in $\text{MoO}_3$ and 34\% in KCuSe~ with inclusion of four-phonon scattering \cite{gokhale2021cross, gokhale2023cross}.

In high thermal conductivity materials such as boron arsenide and graphene, three-phonon scattering is weak due to the large phonon band gaps or selection rules [reason (\roml{2}) above]. For boron arsenide, the large acoustic-optical phonon gap and acoustic bunching forbid three-phonon scattering \cite{lindsay2013a} and the predicted thermal conductivity decrease by more than 37\% at room temperature with inclusion of four-phonon scattering \cite{feng2017four}.  Similarly, in graphene, symmetry selection rules restrict phonon scattering processes involving an odd number of flexural (out-of-plane) phonons \cite{lindsay2010}, enabling high thermal conductivity values of up to 3500 W/m-K when only three-phonon scattering is considered \cite{lindsay2014} and this reduces to less than 900 W/m-K with inclusion of four-phonon scattering \cite{feng2018}.
There are subsequent developments for graphene, based on temperature-dependent potential energy surface by Gu et al.~\cite{gu2019revisiting}, but the predicted thermal conductivity remains significantly low at $\sim$1500 W/m-K compared to the three-phonon counterpart.

It is important to note that if four-phonon scattering is required due to material/condition being strongly anharmonic [reason (\roml{1}) above], then phonon renormalization is also often important, as discussed in Sec.~\ref{sec_renormalization}.

Further, we also note that the quartic force constants, which are needed in Eqn.~\ref{eqn_quartic_IFC}, are quite sensitive to the choice of displacement amount when evaluated using the finite-difference approach. For instance, Feng et al. \cite{zhou2023extreme} have shown that the thermal conductivity of silicon varies from $\sim$20 to $\sim$120 W/m-K with a change in displacement amount from 0.01 to 0.03\textup{~\AA} at 300 K using the LDA functional.

\subsection{Temperature-dependent Potential Energy Surface}
\label{sec_IFC}
In the lowest-level theory, the interatomic force constants are obtained either using the finite-difference approach or by using the force-displacement dataset fitting on randomly displaced structures (using either $L$1-norm, i.e., compressive sensing, or $L$2-norm, i.e., least square error approaches) \cite{esfarjani2008, zhou2014lattice,  eriksson2019}. The IFCs obtained using this approach correspond to zero-temperature sampling of the potential energy surface. At finite temperatures, the atomic displacements around static positions are according to their thermal displacements, and the potential energy surface experienced by atoms with thermal perturbations could be different than that experienced at 0 K. This is particularly the case for materials with weakly bonded atoms \cite{ravichandran2018unified}. 

To account for the temperature-dependence of the potential energy surface, molecular-dynamics simulations can be carried out to obtain the thermal trajectory of atoms \cite{xia2020}. Several non-related thermal snapshots can be sampled from this thermal trajectory for generating an over-specified (more equations than the number of unknown interatomic force constants) force-displacement dataset. This approach, however, requires molecular dynamics simulations for several thousand timesteps using large computational supercells, and the obtained thermal datasets correspond to classical statistics. An alternative approach for obtaining the thermal displacement-force dataset is to first compute the phonon dispersion (phonon spectrum) using the 0 K force constants and then use the obtained phonon spectrum to get thermal displacements of atoms via \cite{west2006, hellman2013, shulumba2017}:
\begin{equation}
    \begin{split}
        \label{eqn_thermalPopulation}
        u_{b,l}^{\alpha} = \frac{1}{\sqrt{N}}
        \sum_{\boldsymbol{q}\nu}
        \sqrt{\frac{\hbar(n_{\boldsymbol{q}\nu} + 1)}
            {m_b\omega_{\boldsymbol{q}\nu}}}
            \cos{(2\pi\eta_{1, \boldsymbol{q}\nu})}
            \times \\
            \sqrt{-\ln{(1-\eta_{2, \boldsymbol{q}\nu})}}
            {{\tilde{e}^\alpha}_{b, \boldsymbol{q}\nu}}
            e^{i \boldsymbol{q}  \cdot{\boldsymbol{r}}_{0l} },
    \end{split}
\end{equation}
where $\eta_{1, \boldsymbol{q}\nu}$ and $\eta_{2, \boldsymbol{q}\nu}$ are random numbers sampled from a uniform distribution and constrained by $\eta_{1, \boldsymbol{q}\nu} = \eta_{1, -\boldsymbol{q}\nu}$ and $\eta_{2, \boldsymbol{q}\nu} = \eta_{2, -\boldsymbol{q}\nu}$. 

 The harmonic and anharmonic force constants can be obtained by fitting Eqn.~\ref{eqn_PE} to the generated force-displacement dataset. Several approaches/choices have been reported/adapted in the literature for this force-displacement dataset fitting:
\begin{itemize}
\item Simultaneous fitting of harmonic and anharmonic force constants:  As the name suggests, in this approach, harmonic and anharmonic force constants are simultaneously fitted to the force-displacement dataset. This approach is utilized, for instance, in Refs.\cite{han2023thermal} for including the effect of temperature on harmonic and anharmonic force constants for studying thermal transport in graphene, and is sometimes referred to as thermal renormalization.
\item Anharmonic force constants fitting on residual-force-displacement dataset obtained by harmonic contribution removal: The phonon thermal conductivity obtained via the lattice dynamics approach is quite sensitive to harmonic force constants. As such, many studies obtain harmonic force constants using more stringent convergence criteria involving much larger supercells/phonon wavevector grids using density functional perturbation theory (corresponding to 0 K) \cite{ravichandran2018unified}. The contribution of these harmonic force constants can be removed from the force-displacement dataset forces and the generated residual forces-displacement dataset can be fitted to obtain the anharmonic force constants. After obtaining anharmonic force constants, the effect of temperature can be included in harmonic force constants via phonon renormalization (as discussed in Sec.~\ref{sec_renormalization}). While extracting anharmonic constants, one has an option of fitting residual forces, first only to cubic force constants, followed by subtraction of the extracted cubic force constant contribution from residual forces and subsequent fitting of leftover forces to quartic force constants, or fitting both cubic and quartic force constants simultaneously. The former is motivated when only three-phonon scattering is employed, so that all anharmonicity is included in cubic force constants and hence thermal conductivity via the three-phonon scattering. The example of studies employing the former approach is Ref.~\cite{klarbring2020}, whereas examples of studies employing the latter approach of simultaneous cubic/quartic force constants fitting are Refs. \cite{ravichandran2018unified, jain2020,gu2019revisiting}.
\end{itemize}

\begin{figure}
\centering
\epsfbox{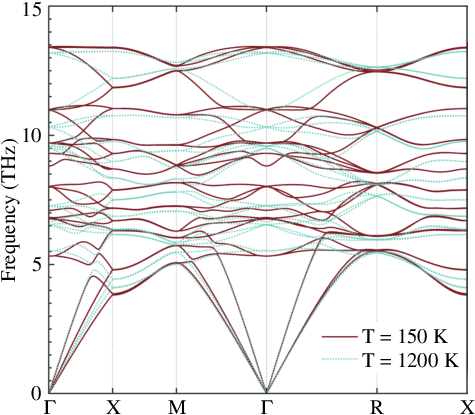}
\caption{\textbf{Temperature-dependent potential energy surface.} The effect of temperature-dependent interatomic force constants on the phonon frequencies of FeSi, with solid and dashed lines representing phonon dispersion at 150 K and 1200 K, respectively. [ Reprinted with permission from Ref. \cite{hellman2013temperature}] }
\label{fig_FD-TDEP}
\end{figure}

\subsection{Phonon Renormalization}
\label{sec_renormalization}
\begin{figure}
\centering
\epsfbox{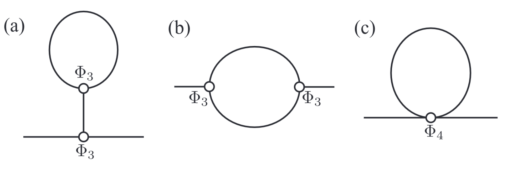}
\caption{\textbf{Feynman diagrams for phonon renormalization.} Feynman diagrams representing (a) Tadpole, (b) Bubble, and (c) Loop shifts. [Reprinted with permission from reference \cite{tadano2018first}] } 
\label{fig_feynman}
\end{figure}

The phonons defined using Eqns.~\ref{eqn_dynamical} and \ref{eqn_eigen} are harmonic quasi-particles. The effect of anharmonicity on phonon frequencies/eigenshapes is included via phonon renormalization, which can be captured via self-energy corrections using the Green’s function formalism \cite{wallace1966a}. Phonon renormalization using Green’s function formalism involves solving the Dyson equation  \cite{tadano2018first} and, at the lowest order, results in the Tadpole, Bubble, and Loop diagrams, as shown in Fig.~\ref{fig_feynman}. The Tadpole and Bubble shifts are the second-order Feynman diagrams obtained from cubic force constants.  The Tadpole shift is present only for materials with non-primitive unitcells and its contribution is small compared to the Bubble shift.  The loop diagram is the first-order contribution from the quartic terms and is significant in materials with strong anharmonicity. The expressions of phonon frequency shifts arising from this approach of renormalization are available in Ref.~\cite{turney2009a, tadano2018first}. The phonon renormalization performed using this approach results only in phonon frequency changes, i.e., the corresponding phonon eigenvectors remain non-renormalized.

An alternative approach employed in literature for phonon renormalization is to renormalize harmonic interatomic force constants  \cite{wallace1966a}:
\begin{widetext}
\begin{eqnarray}
\Phi^{c, \alpha\beta}_{ij} &=& 
    \braket{\frac{\partial^2 U}{\partial u_i^{\alpha} \partial u_j^{\beta}}} \\
    &=&  \Phi^{o, \alpha\beta}_{ij} + \frac{1}{2}
    \sum_{l^{'''}l^{''''}} \sum_{b^{'''}b^{''''}}\sum_{\gamma\delta} \Xi_{ijkl}^{\alpha\beta\gamma\delta} \braket{u_i^{\alpha} u_j^{\beta}} \\
    &=& \Phi^{o, \alpha\beta}_{ij} + 
    \frac{\hbar}{4N}\sum_{l^{'''}l^{''''}} \sum_{b^{'''}b^{''''}}\sum_{\gamma\delta}\sum_{\boldsymbol{q}\nu}
    \Xi_{ijkl}^{\alpha\beta\gamma\delta}
    \frac{{{\tilde{e}^\gamma}_{b^{'''}, \boldsymbol{q}\nu}}
    {{\tilde{e}^{\dagger\delta}}_{b^{''''}, \boldsymbol{q}\nu}}}
    {\omega_{\boldsymbol{q}\nu}\sqrt{m_{b^{'''}} m_{b^{''''}}}}
    (2n_{\boldsymbol{q}\nu}+1)
    e^{i\boldsymbol{q}\cdot{ (\boldsymbol{r}_{0l^{'''}} - \boldsymbol{r}_{0l^{''''}}) }}, \label{eqn_renorm}
\end{eqnarray}
\end{widetext}
where $\Phi^{c, \alpha\beta}_{ij}$, $\Phi^{o, \alpha\beta}_{ij}$ represent corrected, original harmonic force constants and $\braket{.}$ represents thermal averaging. The corrected force constants are obtained by starting with the original harmonic force constants and by iterating through Eqn.~\ref{eqn_renorm} using the original quartic force constants. During iterations, the harmonic properties (phonon frequencies, eigenvectors, occupations, etc.) used in Eqn.~\ref{eqn_renorm} are calculated using the updated harmonic force constants. After convergence of Eqn.~\ref{eqn_renorm}, the cubic and quartic force constants are also renormalized by refitting the residual force-displacement dataset by subtracting corrected harmonic force contributions. 

Please note that the renormalization performed using Eqn.~\ref{eqn_renorm} accounts for the Loop shift. However, since the Bubble shift is second-order in cubic force constants, the contribution of the Bubble shift is often small in many material systems compared to the Loop shift \cite{tadano2018quartic}. Further note that the effect of renormalization is included in cubic force constants indirectly via residual force-displacement dataset fitting \cite{ravichandran2018unified}.

\begin{figure}
\centering
\epsfbox{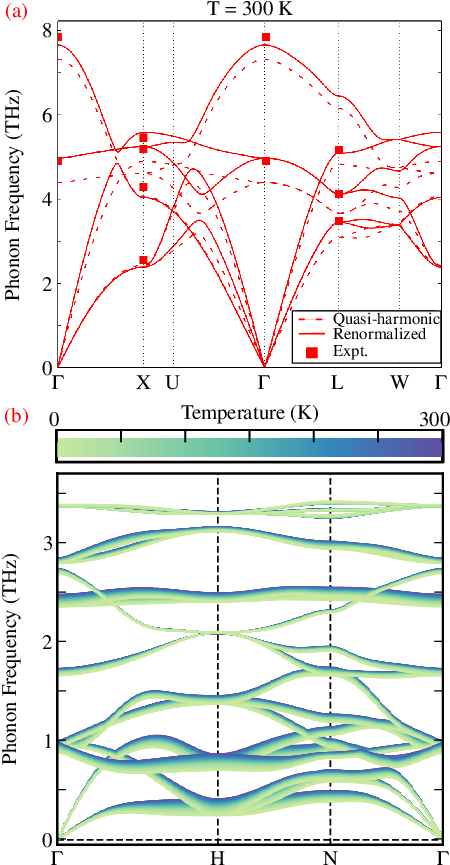}
\caption{\textbf{Phonon Renormalization.} The effect of phonon renormalization on the phonon frequencies of (a) NaCl at 300 K and (b) $\text{Tl}_3\text{VSe}_4$. The experimental datapoints in (a) are from Ref.~\cite{raunio1969}. [Reprinted with permission from Ref. \cite{ravichandran2018unified, jain2020}].}
\label{fig_renorm}
\end{figure}

The examples of studies incorporating phonon renormalization via the Green's function approach are Refs.~\cite{tadano2018quartic, ohnishi2022anharmonic, yao2023anharmonic} while the examples of the latter approach based on harmonic interatomic force constants renormalization include Refs \cite{ravichandran2020,xia2020,jain2020, yue2023strong} (Fig.~\ref{fig_renorm}). The examples of studies employing a hybrid approach where the Loop shift is accounted for self-consistently using the harmonic interatomic force constants renormalization and the Bubble shift are included subsequently on the obtained phonon frequencies include Refs.~\cite{tadano2022first}.

\subsection{Multi-channel Thermal Transport}
\label{sec_}
For low thermal conductivity materials (typically the case when material/conditions are strongly anharmonic), it has been recently reported that the thermal conductivity contribution obtained using the particle-like phonon-only channel (via the BTE approach) is insufficient, and an additional contribution originating from wave-like tunneling of modes (referred to as coherent channel) is also significant \cite{mukhopadhyay2018, simoncelli2019}. The contribution of this coherent channel is negligible for stiff materials, and the thermal transport in such cases is dominated by the particle-channel only. On the other extreme, for weakly-bonded materials with strong anharmonicity or completely disordered materials (such as amorphous solids), the coherent channel is the main contributor to thermal transport \cite{simoncelli2019}. The contribution of the coherent channel towards the lattice thermal conductivity can be obtained using the formulation of Simmoncelli et al.\cite{simoncelli2019} as:
\begin{equation}
\begin{split}
    \label{eqn_coherent}
k^c_{\alpha\beta} 
 = 
 \frac{\hbar^2}{k_BT^2}
 \frac{1}{VN}
 \sum_{\boldsymbol{q}}
 \sum_{(\nu_{} \ne \nu_{1})}
 \frac{ \omega_{\boldsymbol{q}\nu_{}} +  \omega_{\boldsymbol{q}\nu_{1}} }{2}
 V^{\alpha}_{\boldsymbol{q}, \nu\nu_{1}}
 V^{\beta}_{\boldsymbol{q}, \nu_1\nu_{}} 
 \times
 \\
 \frac{
  \omega_{\boldsymbol{q}\nu_{}}  n_{\boldsymbol{q}\nu_{}} (n_{\boldsymbol{q}\nu_{}} + 1)
  +
  \omega_{\boldsymbol{q}\nu_{1}}  n_{\boldsymbol{q}\nu_{1}} (n_{\boldsymbol{q}\nu_{1}} + 1)
 }
 {
 4(\omega_{\boldsymbol{q}\nu_{}} - \omega_{\boldsymbol{q}\nu_1})^2 
 + (\Gamma_{\boldsymbol{q}\nu_{}} + \Gamma_{\boldsymbol{q}\nu_{1}} )^2
 }
 (\Gamma_{\boldsymbol{q}\nu_{}} + \Gamma_{\boldsymbol{q}\nu_{1}}),
\end{split}
\end{equation}
where $\Gamma_{\boldsymbol{q}\nu_{}}$ is phonon linewidth ($\Gamma_{\boldsymbol{q}\nu_{}} = 1/\tau_{\boldsymbol{q}\nu_{}}$) and $V^{\alpha}_{\boldsymbol{q}, \nu\nu_{1}}$ is the $\alpha$-component of velocity operator and can be obtained as:
\begin{equation}
    {\boldsymbol{V}}_{\boldsymbol{q}, \nu\nu_{1}} = 
    \frac{1}{2\sqrt{ \omega_{\boldsymbol{q}\nu_{}} \omega_{\boldsymbol{q}\nu_{1}} }} 
    \bra{ {{{\boldsymbol{e}}}_{\boldsymbol{q}\nu}}} 
    \frac{\partial \boldsymbol{D}_{\boldsymbol{q}} }{\partial \boldsymbol{q}} 
    \ket{ {{{\boldsymbol{e}}}_{\boldsymbol{q}\nu_{1}}}}.
\end{equation}
The total thermal conductivity can then be obtained as:
\begin{equation}
\label{eqn_}
k = k^{p} + k^{c},
\end{equation}
where $k^{p}$ and $k^{c}$ are the contributions of population and coherent channel towards the thermal transport. 

\begin{figure}
\centering
\epsfbox{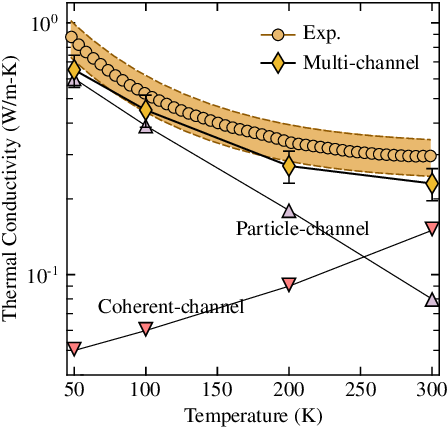}
\caption{\textbf{Mulit-channel thermal transport.} The thermal conductivity of $\text{Tl}_3\text{VSe}_4$ obtained from multi-channel thermal transport theory. The experimental results are from Mukhopadhyay et al.~\cite{mukhopadhyay2018}. [Reprinted with permission from Ref. \cite{jain2020}]. }
\label{fig_coherent}
\end{figure}

The thermal transport described using Eqn.~\ref{eqn_} is referred to as multi-channel thermal transport and is reported to be significant in low-thermal conductivity materials such as   $\text{Tl}_3\text{VSe}_4$\cite{jain2020,mukhopadhyay2018},$\text{Cs}_2\text{AgBiBr}_6$\cite{zheng2024unravelling}, and $\text{Cs}_3\text{Bi}_2\text{I}_6\text{Cl}_3$\cite{acharyya2022glassy}.
The discussion on multi-channel thermal transport originated from the study of Mukhopadhyay et al.~\cite{mukhopadhyay2018}, where the authors noted that the thermal conductivity obtained from particle-like channel falls short of the experimentally measured value by a factor of two or more [see Fig.~\ref{fig_coherent}]. The authors proposed ad-hoc contribution coming from the amorphous channel (subsequently termed as coherent channel), motivated by the failure of the Ioffe-Regel limit. Subsequently, Simoncelli et al.~\cite{simoncelli2019} proposed a unified thermal transport theory applicable for both amorphous/disordered and crystalline materials. Since the contribution of coherent channel is significant in low thermal conductivity strongly anharmonic materials, the corrections from other higher-order theories, such as four-phonon scattering and phonon renormalization, are also important in such cases. For instance, Xia et al.~\cite{xia2020} suggested that the thermal transport in $\text{Tl}_3\text{VSe}_4$ is describable using only particle-like channel when phonon renormalization is also included. However, subsequently, Jain \cite{jain2020} discussed that when all of the higher-order effects are included along with lattice thermal expansion, the particle-like channel still falls short of the experimental value by a factor of two at 300 K, and coherent-channel is needed to describe the experimentally measured results.
Since then, the multi-channel thermal transport is reported in many material systems, for instance,  Li et al.~\cite{li2023wavelike} predicted ultra-low thermal conductivity of only $0.2$ W/m-K in layered perovskite $\text{Cs}_3\text{Bi}_2\text{I}_6\text{Cl}_3$, which is consistent with experimental measurement by Acharyya et al.\cite{acharyya2022glassy} and suggested that the contribution of wave-like tunneling modes is more than 70\% at 300 K.
Similarly, Zheng et al.\cite{zheng2024unravelling} investigated thermal transport in $\text{Cs}_2\text{AgBiBr}_6$ by accounting for both cubic and quartic anharmonic phonon renormalization, as well as the contribution of wave-like phonon tunneling. Their analysis revealed that, when only three-phonon scattering processes are considered, the particle-like transport channel contributes more than 50\% to the total thermal conductivity. However, upon inclusion of four-phonon scattering, the particle-like contribution is significantly suppressed, and the wave-like channel becomes dominant, accounting for over 50\% of the thermal conductivity at temperatures above 310 K.

\begin{figure*}
\centering
\epsfbox{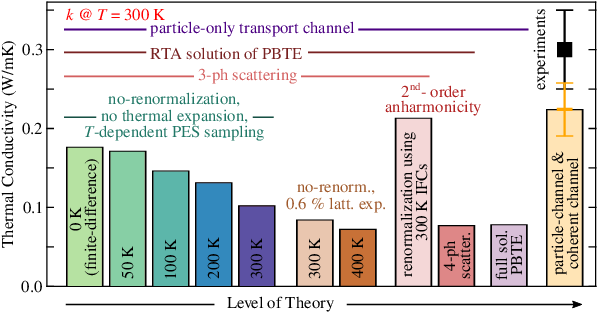}
\caption{\textbf{Higher-order thermal transport effects in $\text{Tl}_3\text{VSe}_4$.} The thermal conductivity of $\text{Tl}_3\text{VSe}_4$ obtained by subsequenctly including higher-order thermal transport physics. The experimental result is from Mukhopadhyay et al.~\cite{mukhopadhyay2018}. [Reprinted with permission from Ref. \cite{jain2020}]. }
\label{fig_multieffect}
\end{figure*}

As mentioned above, the multi-channel thermal transport becomes relevant when the materials/conditions are strongly anharmonic and phonon mean free paths become smaller than the Ioffe-Regel limit. In such scenarios, it is also required to carry out temperature-dependent sampling of the potential energy surface, include both three- and four-phonon scattering, and carry out phonon renormalization. The subsequent role of these higher-order effects on thermal conductivity of $\text{Tl}_3\text{VSe}_4$ is reported in Ref.~\cite{jain2020} and is reprinted here in Fig.~\ref{fig_multieffect}. It is worthwhile to note that while we did not discuss lattice thermal expansion in higher-order effects (quasi-harmonic approximation), it can also become relevant under anharmonic conditions (as shown in Fig.~\ref{fig_multieffect}).

Based on our experience, we recommend the following criteria for selecting higher-order thermal transport physics in different materials:
\begin{itemize}

\item \textbf{Case-1: Materials with high Debye temperature (with respect to temperature of interest)}.
The fraction of phonons undergoing \textit{Normal} scattering becomes significant at low temperatures when only close-to-Gamma point phonons in the Brillouin zone are active. This can be accessed via the Debye temperature. For silicon, the Debye temperature is around 600 K \cite{Keesom1959} and the change in thermal conductivity with full solution of the BTE is less than 10\% at 300 K, whereas for diamond, the Debye temperature is more than 2000 K \cite{tohei2006debye} and the corresponding change is more than 50\% at 300 K. At temperatures much lower than the Debye temperature, the full solution of the BTE is needed.

\item \textbf{Case-2: Materials with quadratic/sub-linear dispersion of phonons (like in two-dimensional materials)}.
If the material has quadratic dispersion near the Gamma-point in the Brillouin zone, then the fraction of phonons with \textit{Normal} scattering is high and the full solution of the BTE is required. 

\item \textbf{Case-3: Three-phonon scattering is weak [either due to symmetry (like in graphene) or due to phonon frequency gaps (in BAs)]}.
In such cases, four-phonon scattering is also required along with the three-phonon scattering.

\item \textbf{Case-4: Material/conditions are extremely anharmonic}.
This is often reflected in large thermal mean square displacements of atoms, short phonon mean free paths (compared to interatomic distance), and low thermal conductivity from the lowest-order theory. In such cases, lattice thermal expansion, temperature-dependent sampling of the potential energy surface, phonon renormalization, four-phonon scattering, and multi-channel thermal transport are all required for the correct description of thermal transport physics.

\end{itemize}

Before ending this Section, we would like to note that the full solution of the BTE is implemented in openly available software packages ShengBTE \cite{ShengBTE_2014}, almaBTE \cite{carrete2017almabte}, and phono3py \cite{phono3py}. The temperature-dependent sampling of the potential energy surface can be carried out using the TDEP package \cite{knoop2024tdep} and ALAMODE \cite{Tadano2014-qa}. The four-phonon scattering is implemented in the package fourPhonon \cite{feng2017four}, and the multi-channel thermal transport physics is implemented in phono3py \cite{simoncelli2022wigner}. Further, multiple tools are available in the literature to obtain interatomic force constants using finite-difference \cite{phono3py, ShengBTE_2014}, compressive sensing \cite{zhou2014lattice}, and machine-learning \cite{eriksson2019}.

\section{Phonon Thermal Conductivity Prediction: Computational Challenges}
\label{sec_challenges}

Over the last decade, there have been many developments in the phonon thermal transport prediction tools, and many of the theoretical functionalities discussed above are now available in openly available computational tools. However, despite the availability of such tools, the computation of interatomic force constants and phonon-phonon scattering rates has remained a major computational bottleneck in high-throughput discovery of materials for thermal applications. The computational challenge associated with interatomic force constants is due to the number of required DFT force evaluations for extraction of cubic and quartic anharmonic force constants whereas the challenge with phonon-phonon scattering rates evaluation is due to summations involved in Eqns.~\ref{eqn_3ph}, \ref{eqn_cubic_Fourier} and Eqns.~\ref{eqn_4ph}, \ref{eqn_quartic_IFC}. For instance, for simple semiconductor Si with two atoms in the primitive unitcell, the number of symmetry-reduced cubic force constants and quartic force constants are 126 and 14 within 2.5 $\text{\AA}$ and 1.5  $\text{\AA}$ cutoff radius (the corresponding numbers are 31482 and 4698 without considering crystal symmetries). For more complex materials, these numbers increase sharply:  1367 and 1033 for BGG clathrate with 7.2 $\text{\AA}$ and 6 $\text{\AA}$ interaction cutoffs; requiring similar number of DFT force evaluations on large computational cells (often 200-400 atoms in each cell) for force constants evaluation via the finite-difference approach. Similarly, the typical number of phonon modes required for converged thermal conductivity prediction is $\sim10^4$, each of which requires $\sim10^8$ evaluations of Eqn.~\ref{eqn_3ph}, with each evaluation, in turn, requiring $\sim10^{7}$ evaluations of Eqn.~\ref{eqn_cubic_Fourier} for three-phonon scattering rates. For four-phonon scattering, $\sim10^{12}$ evaluations of Eqn.~\ref{eqn_4ph} are required with each evaluation requiring $\sim10^{9}$ evaluations of Eqn.~\ref{eqn_quartic_IFC}. Recently, several machine learning based approaches have been presented in the literature to overcome these computational challenges as discussed below.

\subsection{Machine Learning for Interatomic Force Constants}
\label{sec_ML_IFC}
The earlier efforts to reduce the computational cost of interatomic force constants using machine learning approaches were motivated by force-field developments for molecular dynamics simulations  \cite{ bartok2010, lysogorskiy2021performant, Podryabinkin2023, fan2021neuroevolution}. These forcefields were developed by employing DFT data from various pristine, perturbed, strained, and defective configurations  \cite{Cheng2023, Khorshidi2016AmpAM, fan2021neuroevolution}. The data is converted into descriptors representing local chemical environments, and the machine learning models are trained to predict forces on atoms based on these descriptors. Many different descriptors and machine learning models have been employed in the literature for this purpose  \cite{Podryabinkin2023,fan2021neuroevolution,Yao2018,Wang2018, bartok2013,Abbott2019}. Such trained forcefields have been employed to obtain interatomic force constants to study thermal transport in bulk materials using the lattice dynamics approach \cite{ladygin2020lattice,fan2021neuroevolution}. However, the development/training of such machine-learned forcefields requires a large amount of data (for instance, 325 non-correlated configurations of force-displacement data for bulk PbTe \cite{fan2021neuroevolution}).

An alternative approach to obtaining anharmonic interatomic force constants was recently suggested by some of us \cite{Srivastava2024b}. In this approach, the learning is carried out only on the local potential energy surface using targeted machine learning. The harmonic interatomic force constants are obtained from DFT and are used to obtain the thermal perturbation of atoms around their equilibrium positions. After obtaining forces on a few such thermal configurations from DFT, a local/targeted machine-learning model is trained on this data, and this trained model is sufficient to obtain all anharmonic (cubic and quartic) force constants of a given material. We have tested this approach for $\sim$230 materials and obtained a mean absolute percentage error on thermal conductivity to be less than 10\% \cite{Srivastava2024b}. 

\begin{figure}
\centering
\epsfbox{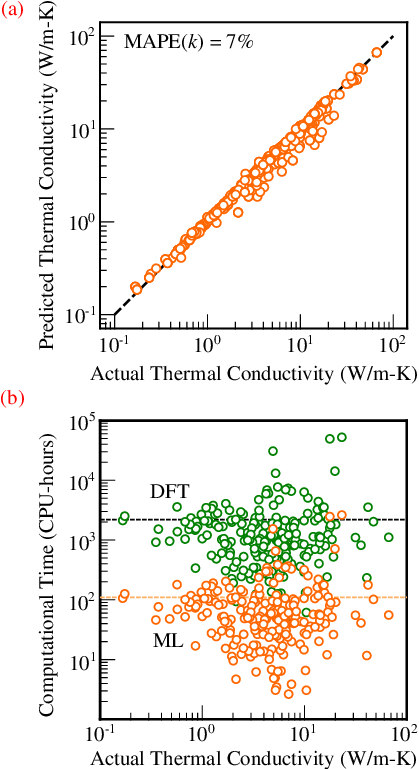}
\caption{\textbf{Machine learning for interatomic force constants.} (a) The thermal conductivity obtained using machine learning based interatomic force constants on 230 diverse materials and its comparison with actual values, and (b) the computational cost of anharmonic interatomic force constants evaluation with and without machine learning. [Reprinted with permission from Ref. \cite{Srivastava2024b}].}
\label{fig_forcefit}
\end{figure}

As universal machine-learned forcefields are now becoming available  \cite{Batatia2022mace,Batatia2022Design,deng2023chgnet}, it is also possible to employ such forcefields for thermal conductivity prediction. For instance, P\'{o}ta et al \cite{Pota-MACE} used MACE-based universal field to predict thermal conductivity of more than 103 binary materials with rocksalt, zincblende, and wurtzite crystal structures. The conductivities obtained from such pre-trained forcefields were within a factor of two of the actual values. Further, on fine-tuning the universal model with new DFT configurations, the prediction error reduced drastically, for instance, from 47\% to 2\% for LiBr with the inclusion of three new DFT configurations.

While these universal forcefields are computationally more expensive to train compared to targeted models, they have to be trained only once. However, these forcefields require proper testing before their use for a given material system/condition, as the material/condition may not be well sampled by the forcefield. For instance, it is possible to use such a forcefield for strained material and get interatomic force constants even if the training is performed only on relaxed, strain-free configurations. However, in such a scenario, the obtained force constants may not be correct.

\subsection{Machine Learning for Phonon-phonon scattering}
\label{sec_ML_phph}
The phenomenal work on the use of machine learning for phonon-phonon scattering rate predictions is reported by Guo et al.~\cite{guo2023}. The authors noted that the phonon lifetimes ($\tau_{\boldsymbol{q}\nu}^{3ph}$, $\tau_{\boldsymbol{q}\nu}^{4ph}$) obtained from Eqns.~\ref{eqn_3ph} and \ref{eqn_4ph}  are  average of millions/billions of phonon linewidths [$\Gamma(\boldsymbol{q}\nu, \boldsymbol{q}'\nu', \boldsymbol{q}''\nu'')$,  $\Gamma(\boldsymbol{q}\nu, \boldsymbol{q}'\nu', \boldsymbol{q}''\nu'', \boldsymbol{q}'''\nu''')$ where $\Gamma(\boldsymbol{q}\nu, \boldsymbol{q}'\nu', \boldsymbol{q}''\nu'')$, $\Gamma(\boldsymbol{q}\nu, \boldsymbol{q}'\nu', \boldsymbol{q}''\nu'', \boldsymbol{q}'''\nu''')$ correspond to right-hand side of Eqns.~\ref{eqn_3ph} and \ref{eqn_4ph} respectively]. Consequently, a machine learning model could be used for each phonon mode by training the model on a fraction of the phonon linewidths. The authors showed that when only 0.3\% of phonon linewidths are employed for machine learning training and the remaining linewidths are obtained from the trained ML model, though the predicted individual linewidths were not accurate, the obtained lifetimes from averaging of predicted linewidths were accurate. The authors further expanded on this in their subsequent work \cite{guo2023b} and showed that, it is possible to remove ML training, and a simple ML predictor, based on the average of training data, is also sufficient for predicting phonon lifetimes (provided dataset size for averaging is statistically sufficient). The authors showed that for four-phonon scatterings in the case of  LiCoO$_2$, this accounts for a reduction in computational cost by a factor of 45, while preserving the accuracy of thermal conductivity prediction to 5\%. 

\begin{figure}
\centering
\epsfbox{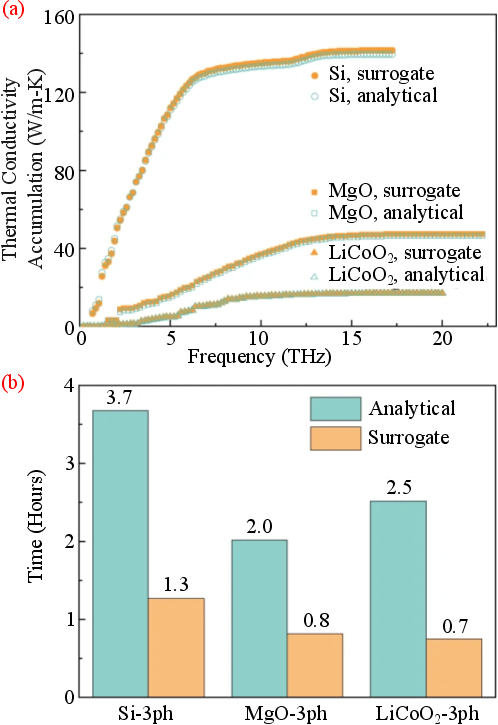}
\caption{\textbf{Machine learning for phonon-phonon scattering rates.} (a) The thermal conductivity accumulation of three diverse materials. The thermal conductivity obtained with both analytical and machine learning (surrogate) models, and (b) the average computational cost saving using the machine learning approach for considered materials [Reprinted with permission from Ref \cite{guo2023}].}
\label{fig_forcefit}
\end{figure}

We also expanded on this work, and presented a complimentary approach, where instead of computing phonon lifetimes for all modes from Eqns.~\ref{eqn_3ph} and \ref{eqn_4ph}, we proposed to use a ML accelerated approach where the ML model is trained on only a fraction of total phonon lifetimes \cite{srivastava2024}. The remaining lifetimes are then obtained from the fitted ML model (as mentioned above, typical thermal conductivity evaluation involves around 10,000 phonon modes). In particular, we have shown that when used in conjunction with the Guo et al.~\cite{guo2023b} approach, our approach results in a reduction of average computational time for phonon-phonon scattering rates evaluation from 70,000 CPU hours to 500 CPU hours for $\sim$230 tested materials \cite{srivastava2024}.

\subsection{End-to-end Machine Learning for direct Thermal Conductivity Prediction}
\label{sec_ML_k}
While the machine learning acceleration approaches discussed above are instrumental in reducing computational cost associated with thermal conductivity prediction, the thermal conductivity evaluation from the lattice dynamics approach still requires significant computations for each material system ($\sim500$ CPU-hours each for force constants and phonon-phonon scattering rate evaluations) even after inclusion of these approaches. As such, research efforts are also focused on direct thermal conductivity prediction from crystal structure (without any mode-level properties), which would require only a few microseconds to predict thermal conductivity directly from structure once the model is trained \cite{Luo2023}. 

In this regard, notable contributions are from Zhu et al.~\cite{zhu2021} where thermal conductivity of diverse material systems is predicted using graph neural network and random forest models and coefficient of determination ($\text{R}^2$) of $0.85$ and $0.87$ are obtained from the two models in a five-fold validation on the dataset consisting of 132 materials. Subsequently, Liu et al.~\cite{liu2022} improved further on this and proposed a transfer learning approach to train a feedforward neural network and obtained an improved $\text{R}^2$ of $0.83$ (compared to $0.67$ with direct learning) on 170 materials in a five-fold cross-validation. More recently, in one of our related works, we have shown that due to limited availability of high-quality systematic thermal conductivity datasets (largest thermal conductivity datasets have $<$1000 diverse enteries\cite{Ojih2023} compared to more than $10^6$ in other material properties datasets \cite{jain2013commentary}), the best performance from such end-to-end models for thermal conductivity prediction is currently limited to 60\% \cite{srivastava2023}.

\section{Conclusions}
\label{sec_conclusion}
The study of phonon-mediated thermal transport in periodic semiconducting solids using the lattice dynamics calculations is reviewed in detail in this paper. The underlying theory for the computation of mode-level phonon properties and thermal conductivity is briefly discussed, starting from the lowest-order theory, which considers the phonon frequencies to be temperature-independent as well as incorporates phonon-phonon scattering only at the three-phonon level to compute phonon scattering rates. While the thermal transport in many intermediate thermal conductivity materials is accurately describable using this level of theory, it has been shown to fail in many high- and low-thermal conductivity materials. 

Over the years, researchers have refined the lattice dynamics approach for thermal conductivity prediction by: (a) improving the accuracy of computed phonon frequencies through temperature-dependent potential energy surface and phonon renormalization, (b) improving the accuracy of computed phonon lifetimes through full solution of the BTE and by including four-phonon scattering, and (c) incorporating multi-channel thermal transport. The underlying theory for each of these higher-order corrections is briefly discussed, along with their prominent success cases. These higher-order phonon calculations are often time-consuming and resource-intensive, and as such, we have provided our recommendations for choosing appropriate higher-order phonon physics for the thermal conductivity calculations in different materials. Implementations of some of these higher-order calculations are available in openly available software packages as discussed in Sec. \ref{sec_limitations}.

The resource-intensive nature of these calculations (especially the large number of DFT force evaluations on large supercells and the computation of phonon-phonon scattering rates) poses a critical challenge for high-throughput calculations for material discovery. Recently, several machine-learning-based approaches have been proposed to mitigate these challenges, and these approaches have demonstrated a significant reduction in the required computational time. These approaches include: (a) machine learning for interatomic force constants, (b) machine learning for phonon-phonon scattering rates evaluation, and (c) end-to-end machine learning for the prediction of thermal conductivity directly from the crystal structure. Together, approaches (a) and (b) have been reported to give more than an order of magnitude reduction in the computational time for thermal conductivity calculation via the lattice dynamics approach.

\section{Declarations}
\label{sec_declaration}

\subsection{Funding} 
The authors acknowledge the financial support from the Core Research Grant, Science \& Engineering Research Board, India (Grant No. CRG/2021/000010), Nano Mission, Government of India (Grant Number: DST/NM/NS/2020/340), and IRCC, IIT Bombay.

\subsection{Conflicts of interests/Competing interests}
The authors have no competing interests to declare that are relevant to the content of this article.

%
\end{document}